\documentclass{article}
\usepackage{amssymb,latexsym,amsmath,amsthm}

\newtheorem{theorem}{Theorem}
\newtheorem{lemma}{Lemma}

\newtheorem{definition}{Definition}
\newtheorem{corollary}{Corollary}

\begin{document}

\newcommand{\bpr}{\operatorname{bpr}}
\newcommand{\indeg}{\operatorname{indeg}}
\newcommand{\pr}{\operatorname{pr}}
\newcommand{\ex}{\operatorname{\mathbf{E}}}
\newcommand{\var}{\operatorname{var}}
\title{Critical Points for Random Boolean Networks}
\author{James F. Lynch\thanks{Research partially supported by NSF
        Grant CCR-9406809}\\
        Department of Mathematics and Computer Science\\
        Box 5815\\
        Clarkson University \\
        Potsdam, NY 13699-5815} 
\date{}
\maketitle

\begin{abstract}
A model of cellular metabolism due to S. Kauffman is analyzed. It consists of
a network of Boolean gates randomly assembled according to a probability
distribution. It is shown that the behavior of the network depends very
critically on certain simple algebraic parameters of the distribution.
In some cases, the analytic results support conclusions based on simulations
of random Boolean networks, but in other cases, they do not.
\end{abstract}
\section{Introduction}
Many dynamical systems are modelled by networks of interacting elements. Examples
come from diverse areas of science and engineering and over enormous scales of
time and space, from biochemical networks within a cell \cite{bb} to food webs
\cite{cn} and collaboration networks in human organizations \cite{wf}.
Often, these systems are subjected to random or unpredictable processes.
In this article, we are concerned with a class
of random networks that S. Kauffman \cite{k.69,k.93} proposed as models of cellular
metabolism. These are networks of Boolean gates, where each gate corresponds to a
gene or protein, and the network describes the interactions among these
chemical compounds. Although Boolean networks capture at least some of the
salient features of the operation of the genome, researchers have been mainly
interested in certain abstract properties of their dynamics. Kauffman's thesis
is that randomly assembled complex systems often exhibit ``spontaneous order,"
that is, even though they are not constructed according to any plan, their
behavior is often stable and robust.

Kauffman considered several measures of order, based on the limit cycle
that the network enters. Since a Boolean network has a finite number of
gates, each of which has two possible states, the network itself
has a finite number of states, and it will eventually return to some state 
it had visited earlier. Since the network operates deterministically, it will
keep repeating this sequence of states, which is called the limit cycle.
Among the measures of order that have been considered are
\begin{enumerate}
\item The number of stable gates---gates that eventually stop changing state.
\item The number of weak gates---gates that can be perturbed without changing
the limit cycle that the network enters.
\item The size of the limit cycle.
\end{enumerate}
The key findings of Kauffman's experiments were that networks constructed
from Boolean gates with more than two inputs were usually disordered in all
three senses: a significant fraction of the gates never stabilized and,
when perturbed, caused the network to enter a different limit cycle, and
the size of the limit cycle was exponential in the number of gates. But
networks constructed from gates with two inputs tended to be ordered in
all three senses, in particular, the average limit cycle size was on the order
of the square root of the number of gates.

These results raise many biological and mathematical questions. From the
viewpoint of biology, a basic issue is whether these Boolean networks capture
the essential features of cellular metabolism. Genes are generally active or
inactive, i.e., transcribing their protein or not,
and the transition between the two states usually happens on a short time
scale. Each gene tends to be directly affected by a small number of proteins.
Thus the Boolean network model seems to be at least a rough approximation of
cellular metabolic networks. Also, genomes are the result of evolution, which
involves random events. However, it would be extremely unlikely that the
simple probability distributions used by Kauffman are realistic. He
studied two kinds of random networks contructed from 2-input gates. In
the first kind, all of the 16 Boolean functions of two arguments are equally
likely to be assigned to a gate. This is certainly a reasonable place to
start, given the lack of knowledge about the actual distribution of
functions in real genomic networks. Two of these 16 functions are constants,
i.e., they ignore their inputs and output only one value. Such gates
exhibit an extreme form of order, and it seemed possible that their
presence was the source of order in networks of 2-input gates. However,
Kauffman also ran simulations of randomly constructed networks without
constant gates, where the remaining 14 two argument functions were equally
likely, and the results were similar to those where all 16 functions were
used.

Kauffman proposed another category of functions as the source of order. He
called these the canalyzing functions. A canalyzing function is a Boolean
function for which there exists some argument and some Boolean value such
that the output of the function is determined if the argument has that value.
For example, the 2-argument OR function $x_1 \vee x_2$ is canalyzing because
if either argument has the value 1, then the value of $x_1 \vee x_2$ is 1.
Fourteen out of the sixteen 2-argument Boolean functions, including the
constant functions, are canalyzing, but this proportion drops rapidly among
Boolean functions with more than two arguments. Thus the hypothesis that
nets with many canalyzing gates tend to be ordered, while those with
few of them do not, is consistent with the experimental results.

All of theses definitions and claims have precise mathematical formulations,
so a natural question is whether the experimental results are supported by
proofs. Interestingly, at about the same time that Kauffman
started investigating random Boolean networks, the mathematical
techniques for dealing with random networks were being developed by
P. Erd\H{o}s and A. R\'enyi \cite{er1,er2} and E. Gilbert \cite{gil},
but it was quite some time before any of these techniques were applied to
the analysis of random Boolean networks. The first proofs of any of
Kauffman's claims appear in an article co-authored by a mathematical
biologist (J. Cohen) and a random graph theorist (T. {\L}uczak) \cite{cl}.

Random graph theory is now a flourishing branch of combinatorics. The most
extensively studied version of random graph is the independent edge model.
In this version, there is a probability $p$ (which may depend on the
number of vertices in the graph) such that for each pair of vertices
independently, there is an undirected edge between them with probability $p$.
Graph theorists have discovered many deep and interesting results about
this kind of random graph, but it does not seem to be a good model of the
random networks studied in biology, communications, and engineering.
A major distinction is that the degree distribution of this kind of graph
is Poisson, but the degree distribution of many real-world networks obeys
a power law. A better model for these situations may be random graphs with a
specified degree distribution, which are considered in recent articles by 
M. Molloy and B. Reed \cite{mr1,mr2}. Some other shortcomings of the standard
version of graph pointed out by M. Newman, S. Strogatz, and D. Watts \cite{nsw}
are that it is undirected and has only one type of vertex. They develop
some techniques for dealing with random directed graphs with vertices of
several types. However, even this model lacks the structure needed to model the
dynamic behavior of networks.

Kauffman's Boolean networks are a further extension of the models in \cite{nsw}
that do include this additional structure. The gates of a Boolean network
are vertices assigned a type corresponding to a Boolean function, and the
directed edges indicate the inputs to each gate. But instead of simply
regarding each vertex as a static entity, we are interested in how the functions
of the gates change the state of the network over time. Our random Boolean
networks are specified by a sequence of probabilities $p_1,p_2,\dots$ whose
sum is 1, where for each gate independently, $p_i$ is the probability that
it is assigned the $i$th Boolean function. Once each gate has been assigned
its function, its indegree is determined by the number of arguments of
the function, and its input gates are chosen at random using the uniform
distribution. Lastly, a random initial state is chosen.

Our main results are simple algebraic conditions, derived from the distribution
$p_1,p_2,\dots$ that imply ordered behavior of the first two kinds
mentioned above: almost all gates stabilize quickly, and almost all gates
can be perturbed without affecting the long-term behavior of the network.
Conversely, if the conditions fail, then the networks do not behave in such
an ordered fashion. Our conditions actually imply forms of ordered behavior
stronger than Kauffman's. That is, the gates stabilize in time on the order
of $\log n$, where $n$ is the number of gates, and the effect of a
perturbation dies out within order $\log n$ steps. Consequently, the failure
of our conditions implies forms of disordered behavior that are weaker than
the negations of Kauffman's.

We then apply our main results to the two classes of 2-input Boolean networks
mentioned above. Here, our analysis verifies some of Kauffman's claims for
networks in the first class, but it casts doubt on similar claims for the
other class.
\section{Definitions}
A Boolean network $B$ is a 3-tuple $\langle V,E,\mathbf{f}
 \rangle$ where $V$ is a
set $\{\,1,\dots,n\,\}$ for some natural number $n$, $E$ is a set of 
directed edges on $V$,
and $\mathbf{f}=(f_1,\dots,f_n)$ is a sequence of Boolean functions
such that for each $v \in V$, the number of arguments of $f_v$ is 
$\indeg(v)$, the
indegree of $v$ in $E$. The interpretation is that $V$ is a collection of
Boolean gates, $E$ describes their interconnections, and $\mathbf{f}$ describes their
operation.

The gates update their states
synchronously at discrete time steps $0,1,\dots$. At any time $t$, each
gate $v$ is in some state $x_v \in \{\, 0,1 \,\}$. Letting $\mathbf{x} =
(x_1,\dots,x_n)$, we say that $B$ is in state $\mathbf{x}$ at time $t$.
Let $\indeg(v) = m$ and $u_1 < u_2 < \dots < u_m$ be the gates
such that $( u_i,v ) \in E$ for $i = 1,\dots,m$. These are referred to as
the \emph{in-gates} of $v$. Then the state of $v$ at time $t+1$ is
$y_v = f_v(x_{u_1},\dots,x_{u_m})$. Letting $\mathbf{y} =
(y_1,\dots,y_n)$, we put $B(\mathbf{x}) = \mathbf{y}$.

The next definitions describe the dynamical properties of Boolean networks
that we will analyze.
\begin{definition}
Let $\mathbf{x} \in \{\, 0,1 \,\}^n$.
\begin{enumerate}
\item For $t = 0,1,\dots$, we put $B^t(\mathbf{x})$ for the state of $B$
at time $t$, given that its state at time 0 is $\mathbf{x}$. That is,
\begin{align}
B^0(\mathbf{x}) & = \mathbf{x} \text{, and} \notag \\
B^{t+1}(\mathbf{x}) & = B(B^t(\mathbf{x})) \text{ for all } t. \notag
\end{align}
We also put $B_v^t(\mathbf{x})$ for $y_v$ where $\mathbf{y} =
B^t(\mathbf{x})$.
\item Gate $v$ stabilizes in $t$ steps on input $\mathbf{x}$
if $B_v^{t^\prime}(\mathbf{x}) =
B_v^{t}(\mathbf{x})$ for 
all $t^\prime \ge t$.
\item For $\mathbf{x} \in \{\, 0,1 \,\}^n$ and $v \in \{\, 1,\dots,n \,\}$,
we put $\mathbf{x}^v$ for the state which is identical to $\mathbf{x}$
except that $x_v^v = 1 - x_v$.
\item Let $u,v \in \{\, 1,\dots,n \,\}$ and $\mathbf{x} \in \{\, 0,1 \,\}^n$.
We say that $v$ affects $u$ at time $t$ on input $\mathbf{x}$ if
$B_u^t(\mathbf{x}) \neq B_u^t(\mathbf{x}^v)$. We put $A^t(v,\mathbf{x}) =
\{\, u \in V : v \text{ affects } u \text{ at time } t
\text{ on input } \mathbf{x} \,\}$.
\item Gate $v$ is $t$-weak on input $\mathbf{x}$ if $A^t(v,\mathbf{x}) =
\emptyset$, i.e., $B^t(\mathbf{x}) = B^t(\mathbf{x}^v)$. Gate $v$ is
$t$-strong on $\mathbf{x}$ if it is not $t$-weak on $\mathbf{x}$.
If $\mathbf{x}$ is understood, we simply say $v$ is $t$-weak or $t$-strong.
\end{enumerate}
\end{definition}
For small intervals of time, the dynamical properties described above
are determined by the ``local" structure of the network. That is, the behavior
of a gate over the interval $0,1,\dots,t$ is determined by the portion of the
network consisting of all gates that can reach the gate by a path in $E$ of
length at most $t$. Similarly, the gates affected by a given gate lie in the
portion consisting of all gates reachable from the gate by such a path. Of course,
for large enough $t$, these portions will be the entire network. The next
definitions capture these notions of locality.
\begin{definition}
\begin{enumerate}
\item For any subset $I \subseteq V$,
\begin{align}
S^0_+(I) &= I \text{, and} \notag \\
S^{t+1}_+(I) &= \{\, u : (v,u) \in E \text{ for some } v \in 
  S^t_+(I) \,\} \text{ for } t \ge 0. \notag
\end{align}
That is, $S^t_+(I)$ is the set of gates at the ends of paths of length $t$
that start in $I$. 
Similarly, $S^t_-(I)$ is the set of gates at the beginnings of paths of length
$t$ that end in $I$.
\item  Then
\begin{align}
N^t_+(I) = \bigcup_{s =0}^t S^s_+(I) \text{, and} \notag \\
N^t_-(I) = \bigcup_{s =0}^t S^s_-(I) \notag
\end{align}
are the out- and in-neighborhoods respectively of $I$ of radius $t$.
\end{enumerate}
We put $S^t_+(v)$ for $S^t_+(\{\, v \,\})$ and similarly for the other
notations.
Thus the state of gate $v$ at time $t$ is determined by the states of the
gates in $S^t_-(v)$ and the functions assigned to the gates in
$N^{t-1}_-(v)$.
\end{definition}

As we will show, for sufficiently small $I$ and $t$, the ``typical"
$N^t_+(I)$ and $N^t_-(I)$ induce a forest on $\langle V,E \rangle$,
i.e., there are no
directed or undirected cycles among their gates. If this is the case for
$N^t_+(v)$, then we can give a simple recursive definition of $A^t(v,
\mathbf{x})$.
\begin{definition}
Let $f(x_1,\dots,x_m)$ be a Boolean function of $m$ arguments, and
$\mathbf{x} = (x_1,\dots,x_m) \in \{\, 0,1 \, \}^m$ be an assignment of 
0's and 1's to its arguments. For $i \in \{\, 1,\dots,m \,\}$, we say that 
argument $i$ directly affects $f$ on input $\mathbf{x}$ if $f(\mathbf{x}) \neq
f(\mathbf{x}^i)$. We extend this notion to gates in a Boolean network in the
obvious way. Given a Boolean network $B$ where gate $v$ has in-gates
$u_1<\dots<u_m$ and state $\mathbf{x} \in \{\, 0,1 \,\}^n$,
for $i = 1,\dots,m$, $u_i$ directly affects $v$ on input $\mathbf{x}$ if
$B_v(\mathbf{x}) \neq
B_v(\mathbf{x}^{u_i})$. 
\end{definition}
\begin{lemma} Assume $N_+^t(v)$ induces a tree on $E$. 
Then for any $s \le t$, any $\mathbf{x} \in \{\, 0,1 \,\}^n$,
and any gate $u \in S_+^s(v)$, $v$ affects $u$
at time $s$ on input $\mathbf{x}$ if and only if
\begin{enumerate}
\item $s = 0$ and $u = v$, or
\item $s > 0$ and, letting $w$ be the unique gate such that
$w \in S_+^{s-1}(v) \cap S_-^1(u)$, $v$ affects $w$ at time $s-1$ on input
$\mathbf{x}$, and $w$ directly affects $u$ on input $B^{s-1}(\mathbf{x})$.
\end{enumerate}
\end{lemma}
\section{Random Boolean Networks} We will be examining randomly constructed Boolean
networks. The random model we use appears to be sufficiently general to
capture  the particular classes of random Boolean networks
in the literature. Let $\phi_1,\phi_2,\dots$ be some ordering of all the finite
Boolean functions, and let $p_1,p_2,\dots$ be a sequence of probabilities such that
$\sum_{i=1}^\infty p_i = 1$. The selection of a random Boolean network with
$n$ gates is a
three stage process. First, each gate is independently assigned a Boolean function
using the distribution $p_1,p_2,\dots$. That is, for each $v = 1,\dots,n$ and
$j = 1,2,\dots$, the probability that gate $v$ is assigned $\phi_j$ is $p_j$.
Next, the in-gates for each gate are selected. If the gate has been assigned an
$m$-argument function, then its in-gates are chosen from the
$\binom{n}{m}$ equally likely possibilities. Finally, a random initial state is
chosen from the $2^n$ equally likely possibilities.

We make several restrictions on the distribution $p_1,p_2,\dots$ still
consistent with the random networks in the literature.
Since we are assuming that
all orderings of the in-gates to a gate are equally likely, for any $j$ and $k$
such that $\phi_j$ and $\phi_k$ are identical except for the ordering of
their arguments, $p_j = p_k$. 
Also, for any $j$ and $k$ such that $\phi_j = \neg\phi_k$, $p_j = p_k$. This
implies that, for any gate $v$ such that $N_-^t(v)$ is acyclic, $B_v^t(\mathbf{x})$
is equally likely to be 0 or 1. Lastly, we
assume that the average and
variance of the number of
arguments of a randomly selected Boolean function, or equivalently, the
average and variance of the indegree of a gate, are finite. That is, letting each 
$\phi_i$ have
$m_i$ arguments, $\sum_{i=1}^\infty p_i m_i^2 \in [0,\infty)$.
\section{Branching Processes}
As will be shown, for $t$ not large compared to $n$, the typical
$N_+^t(v)$ induces a tree in a Boolean network with $n$ gates.
A perturbation of the state of such $v$ may cause perturbations to the states of
$S_+^1(v)$ in the next step, then $S_+^2(v)$, and so on, in a ``wave" that
propagates through $N_+^t(v)$. It is possible that this wave dies out and
the effects of the perturbation are transient, i.e., gate $v$ is weak.
We will show that this behavior
can be approximated by a branching process. Then, by applying basic results
about branching processes, we will derive our results about weak gates.
We will summarize the results that we need. For more information on branching
processes, see T. Harris \cite{har}.

A branching process can be identified with a rooted labelled tree. The tree may
have infinite branches. Each node will be labelled with the unique path from the
root to that node. That is, the root is labelled with the null sequence. If the
root has $k$ children, they are labelled with the sequences $(1)$, $(2)$, \dots,
$(k)$. If the second child of the root has $l$ children, then they are labelled
with the sequences $(2,1)$, $(2,2)$, \dots, $(2,l)$, and so on. Generation $t$
consists of all nodes labelled with a sequence of length $t$. The
number of children of any node is independent of the number of children of any
other node, but the probability of having a certain number of children is the
same for all nodes. Thus the probability space of a branching process is determined
by a sequence $(q_k : k = 0,1,\dots)$ where $q_k$ is the probability that a node
has $k$ children. 
The probability measure on this space will be denoted by $\bpr$. In describing
events in this space, $P$ will denote a branching process. If $\chi$ is a
property of branching processes, $P \models \chi$ means $\chi$ holds for
$P$, and $\bpr(P \models \chi)$ is the probability that $\chi$ holds.

For $t \ge 0$, $P \upharpoonright t$ will be the finite labelled tree which is
$P$ restricted to its first $t$ generations. $Z_t$ will be the random
variable which is the size of generation $t$, i.e., the number of nodes of
depth $t$.

The generating function of the branching process is the series
$$
F(z) = \sum_{k=0}^\infty q_k z^k.
$$
That is, $F(z)$ is the probability generating function of $Z_1$ since
$q_k = \bpr(Z_1 = k)$. A basic result is that the $t$-th iterate of $F(z)$
is the probability generating function of $Z_t$. The iterates of $F(z)$ are
defined by
\begin{align}
F_0(z) &= z \text{ and} \notag\\
F_{t+1}(z) &= F(F_t(z)) \text{ for } t \ge 0. \label{eq.iter}
\end{align}
Then
\begin{theorem}\label{thm.iter}
The probability generating function of $Z_t$ is $F_t(z)$, i.e.,
$$
F_t(z) = \sum_{k=0}^\infty \bpr(Z_t = k) z^k.
$$
\end{theorem}
This enables us to express the moments of $Z_t$ in terms of the moments of
$Z_1$, which in turn have simple representations in terms of the
derivatives of $F(z)$.
Let $\mu$ and $\sigma^2$ be the first and second moments of $Z_1$,
that is, 
$\mu = \ex(Z_1)$ and
$\sigma^2 = \var(Z_1)$.
\begin{theorem}\label{thm.moments}
We have
\begin{align}
\mu &=  F^\prime(1) \text{ and } \notag\\
\sigma^2 &= F^{\prime\prime}(1) + F^\prime(1)
  - (F^\prime(1))^2. \notag
\end{align}
More generally, for all $t \ge 0$, the first and second moments of $Z_t$ are
\begin{align}
\ex(Z_t) &= \mu^t \text{ and } \notag\\
\var(Z_t) &= 
  \begin{cases}
    \dfrac{\sigma^2 \mu^t(\mu^t - 1)}{\mu^2 - \mu} &\text{if $\mu \neq 1$,}\\
    t \sigma^2 &\text{if $\mu = 1$.}
  \end{cases} \notag
\end{align}
\end{theorem}
\section{Weak Gates} \label{sec.weak}
In this section, $\alpha$ and $\beta$ will be positive constants satisfying
$2\alpha\log\delta + 2\beta < 1$ and $\alpha\log\delta < \beta$,
where $\delta = \ex(m_i)$. 
\begin{lemma} \label{lem.tree}
Let $S \subseteq \{\,1,\dots,n\}$, $|S| \le n^\beta$, and $t \le \alpha 
\log n$.
The following events have probability $1 - o(1)$:
\begin{enumerate}
\item For every $v \in S$, $N_-^t(v)$ induces a tree in $\langle V,E \rangle $.
\item For every distinct $u,v \in S$, $N_-^t(u) \cap N_-^t(v) = \emptyset$.
\end{enumerate}
\end{lemma}
\begin{proof}
We show that each of these events fails with probability $o(1)$. The 
calculations are
similar for both events, and we show the work only for event 1.

If 1. fails, then there exist distinct gates $v_1,\dots,v_s$ such that
\begin{description}
\item[] $s \le \alpha \log n$,
\item[] for $i = 1,\dots,s-1$, $v_i$ is an in-gate of $v_{i+1}$, and
\item[] $v_s \in S$,
\end{description}
and distinct gates $w_1,\dots,w_r$ such that
\begin{description}
\item[] $r \le \alpha \log n$,
\item[] for $i = 1,\dots,r-1$, $w_i$ is an in-gate of $w_{i+1}$,
\item[] $w_1 = v_1$, and
\item[] for some $h \in \{\, 1,\dots,s \,\}$, $w_r = v_h$.
\end{description}
Either $h$ above is 1 or greater than 1. The two cases are similar, and we 
will describe only the second. Therefore we can assume $r \ge 2$.
Now $s$, $r$, and $h$ can be chosen in $O((\log n)^3)$ ways.
The gates $v_1,\dots,v_s$ and
$w_2,\dots,w_{r-1}$ can be chosen in $O(n^{s+r-3+\beta})$ ways. For each $j \in \{\,
1,\dots,s-1\,\} - \{\,h-1\,\}$, the probability that $v_j$ is an in-gate of $v_{j+1}$
is 
\begin{align}
\sum_{i=1}^\infty p_i \frac{\binom{n-1}{m_i-1}}{\binom{n}{m_i}} &=
  \sum_{i=1}^\infty p_i \frac{m_i}{n} \notag \\
  &= \frac{\delta}{n} . \notag
\end{align}
Similarly, the probability that each $w_j$ is an in-gate of $w_{j+1}$
for $j = 1,\dots,r-2$ is $\delta/n$.
The probability that both $v_{h-1}$ and $w_{r-1}$ are in-gates of $v_h$ is
\begin{align}
\sum_{i=1}^\infty p_i\frac{\binom{n-2}{m_i-2}}{\binom{n}{m_i}} &=
  \sum_{i=1}^\infty p_i\frac{m_i(m_i-1)}{n(n-1)} \notag\\
  &= O(n^{-2}). \notag
\end{align}
Altogether, the probability that 1. fails is
\begin{align}
O\big((\log n)^3 \times n^{s+r-3+\beta} \times (\delta/n)^{s+r-4}
 \times n^{-2}\big) &= 
  O\big((\log)^3 \delta^{2\alpha\log n} n^{\beta-1}\big) \notag\\
  &= O\big(((\log n)^3 n^{2\alpha\log\delta+\beta-1}\big) \notag\\
  &= o(1). \notag
\end{align}
\end{proof}
We will use the branching process defined as follows. For each $i = 1,2,\dots$ let
$\phi_i$ have $m_i$ arguments, and define
$$
\lambda = \sum_{i=1}^\infty p_i \sum_{j=1}^{m_i}
 \frac{|\{\, \mathbf{x} \in \{\, 0,1 \,\}^{m_i} :
  \text{ argument } j \text{ directly affects } \phi_i \text{ on input } 
\mathbf{x} \,\}|}
  {2^{m_i}}.
$$
Thus $\lambda$ may be regarded as the average number of arguments that directly affect
a random Boolean function with a random input. Since we are assuming all
orderings of the arguments of a Boolean function are equally likely, we can
simplify the definition of $\lambda$ to
\begin{equation}\label{eq.lambda}
\lambda = \sum_{i=1}^\infty p_i m_i
 \frac{|\{\, \mathbf{x} \in \{\, 0,1 \,\}^{m_i} :
  \text{ argument 1 directly affects } \phi_i \text{ on input } 
\mathbf{x} \,\}|}
  {2^{m_i}}.
\end{equation}
The branching process is defined by
$$
q_k = \frac{\lambda^k}{k!} e^{-\lambda}
$$
for $k = 0,1,\dots$. Therefore $F(z) = e^{\lambda z - \lambda}$.
From Theorem \ref{thm.moments},
\begin{align}
\mu &= \lambda, \notag\\
\sigma^2 &= \lambda, \notag\\
\ex(Z_t) &= \lambda^t \text{, and}\notag\\
\var(Z_t) &= 
  \begin{cases}
    \dfrac{\lambda^{t}(\lambda^t - 1)}{\lambda - 1} &\text{if $\mu \neq 1$,}\\
    t \lambda &\text{if $\mu = 1$.}
  \end{cases} \notag
\end{align}
\begin{definition}
Let $T$ be a labelled tree of height $t$, $B = \langle V,E,\mathbf{f} \rangle$ be a Boolean
network, and $\mathbf{x} \in \{\, 0,1 \,\}^n$ be its state. For
$v \in \{\, 1,\dots,n \,\}$, we put $T \Longrightarrow
v$ if
\begin{description}
\item[] $N_-^t(A^t(v,\mathbf{x}))$ induces a tree in $\langle V,E \rangle$, and
\item[] there is an isomorphism from $T$ onto $\langle A^t(v,\mathbf{x}),E \rangle$.
\end{description}
\end{definition}
\begin{lemma}
If $|T| \le n^\beta$ and the height of $T$ is $t \le \alpha\log n$,
then for all $\mathbf{x} \in \{\,0,1\,\}^n$,
$\pr(T \Longrightarrow v) = 
\bpr(P \upharpoonright t \cong T)(1 +o(1))$.
\end{lemma}
\begin{proof}
By Lemma \ref{lem.tree}, if there is an isomorphism $\tau$ from $T$ onto
$\langle A^t(v,\mathbf{x}),E \rangle$, then almost surely $N_-^t(A^t(v,\mathbf{x}))$
induces a tree in $\langle V,E \rangle$. Thus we need only analyze the probability
that $\tau$ exists.
Let $u_1,\dots,u_s$
be the non-leaf nodes of $T$, in lexicographic order.
The construction of $\tau$ is recursive and proceeds
in stages $1,\dots,s$. At each stage $j$, $\tau(u_j)$ has been defined at some previous
stage, and it is extended to the
children of $u_j$. (At stage 1, $\tau(u_1) = v$ has already been defined.)
Also, the Boolean functions assigned to these children are selected.

Thus, assume that at stage $j$, $\tau(u_1),\dots,\tau(u_{K_j})$ have already been
defined, where $j \le K_j$. Let $u_j$ have $k_j$ children. Then there are 
$\binom{n-K_j}{k_j}$
ways of selecting the children of $\tau(u_j)$ in $A^t(v,\mathbf{x})$.
Having chosen these children, we next assign Boolean functions to them.
Independently, for each child $w$ of $\tau(u_j)$, let $\phi_i$ be
assigned to it. This event has probability $p_i$, and the probability that
$\tau(u_j)$ is an in-gate of $w$ is
$$
\dfrac{\binom{n-1}{m_i-1}}{\binom{n}{m_i}} = \frac{m_i}{n}.
$$
Summing over all $i$, we get the probability that $\tau(u_j)$ directly affects
$w$:
$$
\sum_{i = 1}^\infty \frac{p_i m_i \times
 |\{\, \mathbf{x} \in \{\, 0,1 \,\}^{m_i} :
  \text{ argument 1 directly affects } \phi_i \text{ on input } 
\mathbf{x} \,\}|}
  {n2^{m_i}} = \frac{\lambda}n.
$$
Therefore the probability that these $k_j$ gates are directly
affected by $\tau(u_j)$ is $(\lambda/n)^{k_j}$.

Since the events of assigning Boolean functions to all the gates are
independent, the probability that the selected gates belong to $A^t(v,\mathbf{x})$ is
\begin{align}
\prod_{j=1}^s \binom{n - K_j}{k_j} \left(\frac{\lambda}{n} \right)^{k_j} &=
  \left(\prod_{j=1}^s \frac{\lambda^{k_j}}{k_j!}\right) 
  \left(1 - \frac{O(n^\beta)}{n} \right)
  ^{O(n^\beta)} \notag\\
  &= \left( \prod_{j=1}^s \frac{\lambda^{k_j}}{k_j!} \right)
  ( 1 + O(n^{2\beta-1})). \notag
\end{align}
The probability that no other gates are in $A^t(v,\mathbf{x})$ is 
$$
\left( 1 - \frac{\lambda s}{n} \right)^{n - |T|} = 
  e^{-\lambda s} (1 + O(n^{2\beta - 1})).
$$
Therefore
\begin{align}
\pr(T \Longrightarrow v) &= \left( \prod_{j=1}^s \frac{e^{-\lambda}\lambda^{k_j}}
  {k_j!} \right)(1 + o(1)) \notag\\
  &= \bpr(P \upharpoonright t \cong T)(1 + o(1)). \notag
\end{align}
\end{proof}
We say that a property $\chi$ of branching processes depends only on the first
$t$ generations if, for any two branching processes $P_1$ and $P_2$ such
that $P_1 \upharpoonright t \cong P_2 \upharpoonright t$,
either $P_1 \models \chi$ and $P_2 \models \chi$, or $P_1 \nvDash \chi$ and
$P_2 \nvDash \chi$. Thus $\chi$ can be
identified with a set of labelled trees of depth at most $t$. We will also
use the notation $\langle A^t(v,\mathbf{x}),E \rangle \models \chi$ to mean
$\langle A^t(v,\mathbf{x}),E \rangle$ induces a tree in $\langle V,E \rangle$ whose
corresponding branching process satisfies $\chi$.
\begin{theorem} \label{thm.prbpr}
Let $\chi$ be a property of branching processes that depends only on the first
$\alpha \log n$ generations. Then for all $\mathbf{x} \in \{\,0,1\,\}^n$
$$
\pr(\langle A^t(v,\mathbf{x}),E \rangle \models \chi) = \bpr(P \models \chi) + o(1).
$$
\end{theorem}
\begin{proof}
By the previous lemma, it suffices to show that
$\bpr(|P \upharpoonright \alpha \log n| \ge n^\beta) = o(1)$.

If $|P \upharpoonright \alpha \log n | \ge n^\beta$, then
$Z_t \ge n^\beta/(\alpha \log n)$ for some $t = 1,\dots,\alpha \log n$.
Since $\ex(Z_t) = \lambda^t \le \delta^t \le n^{\alpha\log\delta} <
n^\beta/(\alpha\log n$,
\begin{align}
\pr(Z_t \ge n^\beta/(\alpha \log n)) &\le \frac{\var(Z_t)}
  {\big(n^\beta/(\alpha\log n) - \ex(Z_t)\big)^2} \text{ by Chebyshev's inequality} \notag \\
  &= 
  \begin{cases}
    \dfrac{\lambda^{2t-1} + \lambda^{2t-2} + \dots + \lambda^{t}}
         {\big(n^\beta/(\alpha\log n) - \lambda^t\big)^2} \text{ if } \lambda \neq 1 \\
    \dfrac{t\lambda}{\big(n^\beta/(\alpha\log n) - \lambda^t\big)^2} \text{ if } \lambda = 1
  \end{cases} \notag\\
  &= o(1/\log n) \text{ in either case.} \notag
\end{align}
\end{proof}
A gate $v$ such that $N_-^{\alpha\log n}(A^{\alpha\log n}(v,\mathbf{x}))$ is acyclic is
$\alpha\log n$-weak if and only if its corresponding branching process is
extinct within $\alpha\log n$ generations. Clearly this depends only on
the first $\alpha\log n$ generations, so Theorem \ref{thm.prbpr} applies.
By basic results from branching process theory, the probability of extinction
in $t$ generations is $\bpr(Z_t = 0) = F_t(0)$, and $\lim_{t \to \infty}
F_t(0) = r$, where $r$ is the smallest nonnegative root of $z = F(z)$.
Further, when $\mu \le 1$, $r=1$, and when $\mu > 1$, $r < 1$.
Therefore
\begin{theorem} \label{thm.weak}
There is a constant $r$ such that for all $\mathbf{x} \in \{\,0,1\,\}^n$
$$
\lim_{n \to \infty}\pr(v \text{ is } \alpha \log n\text{-weak }) = r.
$$
When $\lambda \le 1$, $r = 1$, and when $\lambda > 1$, $r < 1$.
\end{theorem}
\begin{corollary}
The expected number of $\alpha \log n$-weak
gates in a random Boolean network is asymptotic to $rn$.
\end{corollary}
A stronger result is
\begin{corollary}
The number of $\alpha \log n$-weak
gates in almost all Boolean networks is asymptotic to $rn$.
\end{corollary}
That is, there is a function $\varepsilon(n)$ such that $\varepsilon(n) \to 0$
and, letting the random variable $X_n$ be the number of $\alpha\log n$-weak gates in
a random Boolean network with $n$ gates,
$$
\lim_{n \to \infty} \pr(|X_n - rn| \le n\varepsilon(n)) = 1.
$$
\begin{proof}
By the previous corollary,
$$
\ex(X_n) = rn + n\varepsilon(n),
$$
where $\varepsilon(n)$ is a function such that $\lim_{n \to \infty} 
\varepsilon(n) = 0$.
When $\lambda \le 1$, $r = 1$, so, letting the random variable $Y_n =
n - X_n$,  by  Markov's inequality
$$
\pr\big( Y_n \ge n\sqrt{\varepsilon(n)}\big) = O\big(\sqrt{\varepsilon(n)}\big).
$$
Therefore the corollary holds for $\lambda \le 1$.

When $\lambda > 1$, $r < 1$, and we need to estimate $\var(X_n)$.
Using methods similar to
those in the proofs of Lemma \ref{lem.tree} and Theorems \ref{thm.prbpr}
and \ref{thm.weak}
it can be shown that, for any two distinct gates $u$ and $v$, almost surely
$N_-^{\alpha\log n}(A^{\alpha\log n}(u,\mathbf{x}))$
and $N_-^{\alpha\log n}(A^{\alpha\log n}(v,\mathbf{x}))$ are acyclic,
their intersection is empty, and
$$
\lim_{n \to \infty}\pr(u \text{ and } v \text{ are } \alpha \log n\text{-weak }) =
  r^2.
$$
Therefore
$$
\var(X_n) = r(1-r)n + n^2 \varepsilon^\prime(n)
$$
for some function $\varepsilon^\prime(n) \to 0$.
By Chebyshev's inequality
\begin{align}
\pr(|X_n - rn - n\varepsilon(n)| > n \sqrt[4]{\varepsilon^\prime(n)}) &\le 
  \frac{r(1-r)n + n^2 \varepsilon^\prime(n)}{n^2 \sqrt{\varepsilon^\prime(n)}} \notag\\
  &\to 0, \notag
\end{align}
and the corollary also holds for $\lambda > 1$.
\end{proof}

When $\lambda > 1$, it is also true that most of the $\alpha\log n$-strong gates
affect many other gates when perturbed.
\begin{corollary}
Let $\lambda > 1$.
For almost all random Boolean networks, if gate $v$ is $\alpha\log n$-strong, then
there is a positive $W$ such that for $t \le \alpha\log n$, the number of gates
affected by $v$ at time $t$ is asymptotic to $W \lambda^t$.
\end{corollary}
\begin{proof}
For $t \ge 0$, let $W_t = Z_t/\mu^t$ ($=Z_t/\lambda^t$ in our case).
Again by basic results from branching process theory,
there is a random variable $W$ such that
\begin{align}
\bpr(\lim_{t \to \infty} W_t = W) &= 1 \text{ and} \notag\\
\lim_{t \to \infty} \bpr(Z_t \neq 0 \text{ and } W = 0) &= 0.
\end{align}
From this the corollary follows.
\end{proof}
\section{Forced Gates}
Instead of analyzing the stable gates in a Boolean network, we will study
the forced gates. Since a gate stabilizes if it is forced, this is a
stronger condition, but it seems to be more amenable to combinatorial
analysis.

For the remainder of this section, $t$ will represent a natural number in
the range $0,\dots,\alpha \log n$, and $y$ will be a variable taking
on the values 0 and 1.
Given a Boolean function $\phi(x_1,\dots,x_m)$
and $\mathbf{x} = (x_1,\dots,x_m) \in \{\, 0,1,* \,\}^m$,
we say that $\mathbf{x}$ forces $\phi$ to $y$ if, 
for all $\mathbf{x}^\prime \in
\{\, 0,1 \,\}^m$ such that $x_i = x_i^\prime$ whenever $x_i \neq *$,
$\phi(\mathbf{x}^\prime) = y$. The $*$'s are ``don't care" values, meaning
their value does not affect the value of $\phi$ whenever the remaining
arguments agree with $\mathbf{x}$.
For example, $\phi$ is forced by every $\mathbf{x} \in \{\, 0,1 \,\}^m$;
if $\phi$ is a constant function, then it is forced by every $\mathbf{x} \in
\{\, 0,1,* \,\}^m$; if $\phi(x_1,x_2) =
x_1 \vee x_2$, then it is forced to 0 by $(0,0)$ and to 1 by $(0,1)$,
$(1,0)$, $(1,1)$, $(1,*)$, and $(*,1)$.
We can now give a recursive definition of forcing for the gates of a
Boolean network.
\begin{definition}
A gate $v$ is forced to $y$ in 0 steps if $f_v$ is the
constant function $y$.

For $t \ge 0$, $v$ is forced to $y$ in $t+1$ steps if,
letting $u_1,\dots,u_m$
be its in-gates, there is $\mathbf{x} \in \{\, 0,1,* \,\}^m$ such that
$\mathbf{x}$ forces $f_v$ to $y$ and for each $i = 1,\dots,m$
such that $x_i \neq *$, $f_{u_i}$ is forced to $x_i$ in $t$ steps.
We say that $v$ is forced (in some number of steps) if it is forced to 0 or 1.
\end{definition}
It is clear that forcing is a stronger condition than stability.
\begin{lemma}
If a gate in a Boolean network is forced to $y$ in $t$ steps,
then it stabilizes to $y$ in $t$ steps.
\end{lemma}
Further, conditioning on the event that $N_-^t(v)$ induces a tree,
the probabilities that the
in-gates of $v$ are forced in $t-1$ steps are independent, and there is a 
recursive formula for computing the probability that $v$ is forced in $t$ steps.
Since $N_-^t(v)$ is almost surely a tree for the values of $t$ being considered
here, the conditional probability given by the recursive formula will be
asymptotic to the unconditional probability of being forced in $t$ steps. 

For any natural number $m$ and $\mathbf{x} \in \{\, 0,1,* \,\}^m$, let
$|\mathbf{x}|_0$ be the number of coordinates of $\mathbf{x}$ that are 0, and
similarly for $|\mathbf{x}|_1$ and $|\mathbf{x}|_*$.
For $i = 1,2,\dots$ let $P_i^y(z_0,z_1)$ be the polynomial
in $z_0$ and $z_1$ defined by
$$
P_i^y(z_0,z_1) = \sum_{\substack{\mathbf{x} \in \{\, 0,1,* \,\}^{m_i} \\
  \mathbf{x} \text{ forces } \phi_i \text{ to } y}}
  z_0^{|\mathbf{x}|_0} z_1^{|\mathbf{x}|_1} (1 - z_0 - z_1 )^{|\mathbf{x}|_*}.
$$
Let
\begin{equation}\label{eq.Gy}
G^y(z_0,z_1) = \sum_{i=1}^\infty p_i P_i^y(z_0,z_1).
\end{equation}
Recursively, define
\begin{align}
G_1^y(z_0,z_1) &= G^y(z_0,z_1), \text{ and for } t \ge 1\notag\\
G_{t+1}^y(z_0,z_1) &= G^y(G_t^0(z_0,z_1),G_t^1(z_0,z_1)). \notag
\end{align}
\begin{lemma}
If $N_-^t(v)$ induces a tree, then the probability that $v$ is forced to
$y$ in $t$ steps is $G_{t+1}^y(0,0)$. 
\end{lemma}
From the definition of $G^y$ and the symmetry condition $p_i = p_j$
whenever $\phi_i = \neg\phi_j$, we have $G^0(a,b) = G^1(a,b)$ for all $a$ and $b$,
and therefore $G_t^0(0,0) = G_t^1(0,0)$ for all $t \ge 1$. Therefore letting
\begin{equation}\label{eq.G}
G(z) = 2 G^0(z/2,z/2)
\end{equation}
and defining
\begin{align}
G_1(z) &= G(z), \text{ and for } t \ge 1\notag\\
G_{t+1}(z) &= G(G_t(z)), \notag
\end{align}
\begin{lemma}
If $N_-^t(v)$ induces a tree, then the probability that $v$ is forced 
in $t$ steps is $G_{t+1}(0)$. 
\end{lemma}
\begin{theorem}\label{thm.forced}
There exists $g \in [0,1]$ such that
$$
\lim_{n \to \infty} \pr(v \text{ is forced in } \alpha\log n \text{ steps }) = g.
$$
Further,
$$
\lim_{t \to \infty} G_t(0) = g,
$$
and $g$ is a root of the equation
$$
g = G(g).
$$
\end{theorem}
\begin{proof}
Since
\begin{align}
P_i^0(a,b) + P_i^1(a,b) &\le 
 \sum_{\mathbf{x} \in \{\, 0,1,* \,\}^{m_i}} 
  a^{|\mathbf{x}|_0} b^{|\mathbf{x}|_1} (1 - a - b )^{|\mathbf{x}|_*} \notag\\
  &\phantom{\le}\text{ for nonnegative $a$ and $b$ such that $a + b \le 1$} \notag\\
&= 1, \notag\\ 
2 G(a) &\le \sum_{i=1}^\infty p_i \notag\\
&= 1 \text{ for } a \le 1. \notag
\end{align}
This implies that $G(z)$ is a continuous function on $[0,1]$ and all $G_t(0)$
are bounded above by 1.
We will show that $G_t(0)$ is a strictly increasing
sequence in $t$. Then, 
taking $g =  \sup(G_t(0) : t \ge 1)$, the theorem follows.

To show $G_t(0) < G_{t+1}(0)$, again assuming that
$N_-^{\alpha\log n}(v)$ is a tree, note that the event that
$v$ is forced to $y$ in $t$ steps is characterized by a collection $\mathcal C$
of rooted trees of height at most $t$ whose nodes are labelled with Boolean
functions. Each of these trees is contained in the collection $\mathcal D$ of rooted
labelled trees that characterizes the event that $v$ is forced to $y$ in
$t+1$ steps. Further, some of these trees in $\mathcal C$ are of height $t$,
and their only leaves that are labelled with constant functions have depth $t$.
Take any such tree and replace each leaf that is labelled with a constant
with a subtree consisting of a node labelled with a nonconstant function
and new in-gates all labelled with constants such that the state of the leaf
remains unchanged. The new tree belongs to $\mathcal D$
but not $\mathcal C$ because $v$ will be forced in $t+1$ steps but not $t$ steps.
Therefore $\mathcal D$ is strictly larger than $\mathcal C$, and 
$G_t(0) < G_{t+1}(0)$.
\end{proof}
\begin{corollary}
The expected number of gates that are forced in $\alpha \log n$ steps
is asymptotic to $gn$.
\end{corollary}
\begin{corollary}
The number of gates that are forced in $\alpha \log n$ steps in
almost all Boolean networks is asymptotic to $gn$.
\end{corollary}
\section{Networks of 2-Input Gates}
We now apply the general results of the previous two sections to some
networks studied by Kauffman. As mentioned in the Introduction, he suggested
that
networks with a large proportion of canalyzing gates tend to be stable with
high probability. A Boolean function $f(x_1,\dots,x_m)$ is canalyzing
if it is forced by some $\mathbf{x} \in \{\,0,1,*\,\}^m$ where $x_i \neq *$ for exactly
one $i \in \{\,1,\dots,m\,\}$. Kauffman's claim seems to be supported by
experiments indicating that networks constructed from 2-argument Boolean
functions usually exhibit stable behavior, while those constructed from Boolean
functions with more than 2 arguments do not. Fourteen out of sixteen 2-argument
Boolean functions are canalyzing, but this proportion drops rapidly among
Boolean functions with more than two arguments. However, our analysis
does not support the experimental findings. To explain these results, we
classify the 2-argument Boolean functions into three categories.
\begin{description}
\item[I.] The two constant functions:
\begin{tabbing}
$f(x_1,x_2) = \neg x_2$\=and\=$f(x_1,x_2) = \neg x_2$\kill
$f(x_1,x_2) = 0$\>and\>$f(x_1,x_2) = 1$
\end{tabbing}
\item[II.] The twelve nonconstant canalyzing functions, consisting of
\begin{description}
\item[A.] The four functions that depend on one argument:
\begin{tabbing}
$f(x_1,x_2) = \neg x_2$\=and \=$f(x_1,x_2) = \neg x_2$\kill
$f(x_1,x_2) = x_1$\>and\>$f(x_1,x_2) = \neg x_1$\\
$f(x_1,x_2) = x_2$\>and\>$f(x_1,x_2) = \neg x_2$
\end{tabbing}
\item[B.] The eight canalyzing functions that depend on both arguments:
\begin{tabbing}
$f(x_1,x_2) = \neg x_2$\=and \=$f(x_1,x_2) = \neg x_2$\kill
$x_1 \vee x_2$ \> and \> $\neg x_1 \wedge \neg x_2$ \\
$\neg x_1 \vee x_2$ \> and \> $x_1 \wedge \neg x_2$ \\
$x_1 \vee \neg x_2$ \> and \> $\neg x_1 \wedge x_2$ \\
$\neg x_1 \vee \neg x_2$ \> and \> $x_1 \wedge x_2$ \\
\end{tabbing}
\end{description}
\item[III.] The two noncanalyzing functions exculsive or and equivalence:
\begin{tabbing}
$f(x_1,x_2) = \neg x_2$\=and \=$f(x_1,x_2) = x_2$\kill
$x_1 \oplus x_2$ \> and \> $x_1 \equiv x_2$ \\
\end{tabbing}
\end{description}
Note that each function is paired with its negation. Let $a$, $b$, and $c$ be
the respective sums of the probabilities of the functions of type
I, II, and III, i.e., $a$ is the probability that a gate is assigned a
function of type I, and so on.
We can now express the $\lambda$ parameter of Section \ref{sec.weak}
(see Equation (\ref{eq.lambda})) in terms of $a$, $b$, and $c$.
Clearly, if $\phi_i$ is of type I,
$$
 |\{\, \mathbf{x} \in \{\, 0,1 \,\}^2 :
  \text{ argument 1 directly affects } \phi_i \text{ on input } 
\mathbf{x} \,\}| = 0.
$$
If $\phi_i$ is of type II.A., say $\phi_i(x_1,x_2) = x_1$, then
$$
 |\{\, \mathbf{x} \in \{\, 0,1 \,\}^2 :
  \text{ argument 1 directly affects } \phi_i \text{ on input } 
\mathbf{x} \,\}| = 4,
$$
whereas if $\phi_i(x_1,x_2) = x_2$, then
$$
 |\{\, \mathbf{x} \in \{\, 0,1 \,\}^2 :
  \text{ argument 1 directly affects } \phi_i \text{ on input } 
\mathbf{x} \,\}| = 0.
$$
If $\phi_i$ is of type II.B., say $\phi_i(x_1,x_2) = x_1 \vee x_2$, then
$$
 |\{\, \mathbf{x} \in \{\, 0,1 \,\}^2 :
  \text{ argument 1 directly affects } \phi_i \text{ on input } 
\mathbf{x} \,\}| = 2.
$$
Altogether, the type II functions contribute $b$ to $\lambda$. Lastly, it is
easily seen that if $\phi_i$ is a type III function, then
$$
 |\{\, \mathbf{x} \in \{\, 0,1 \,\}^2 :
  \text{ argument 1 directly affects } \phi_i \text{ on input } 
\mathbf{x} \,\}| = 4,
$$
and therefore the type III functions contribute $2c$ to $\lambda$, giving
$$
\lambda = b + 2c.
$$

To analyze the fixed gates, note that $G(z)$ (see Equations (\ref{eq.Gy})
and (\ref{eq.G}))
is a weighted sum of the 16 terms $2 P_i^0(z/2,z/2)$ corresponding to the
2-argument Boolean functions. This sum can be simplified by using the above
classification and pairing of these functions.

If $\phi_i$ is the constant function $\phi_i(x_1,x_2) = 0$, then $P_i^0(
z/2,z/2) = 1$, but if it is the constant function $\phi_i(x_1,x_2) = 1$,
then $P_i^0(z/2,z/2) = 0$. Therefore the type I functions contribute the
term $a$ to $G(z)$.

If $\phi_i$ is a type II.A. function, say $\phi_i(x_1,x_2) = x_1$, then
$P_i^0(z/2,z/2) = z/2$. If $\phi_i(x_1,x_2) = \neg x_1$, then
$P_i^0(z/2,z/2) = z/2$ again.
If $\phi_i(x_1,x_2)$ is a type II.B. function, say $x_1 \vee x_2$, then
$P_i^0(z/2,z/2) = z^2/4$. If it is $\neg x_1 \wedge \neg x_2$, then
$P_i^0(z/2,z/2) = z - z^2/4$. Altogether the type II functions contribute the
term $bz$ to $G(z)$.

It is easily seen that the two noncanalyzing functions each have
$P_i^0(z/2,z/2) = z^2/2$, and therefore $G(z) = a + bz + cz^2$. The roots of
the equation
\begin{equation}\label{eq.roots}
z = a + bz + cz^2
\end{equation}
are 1 and $a/c$. Since $G(z)$ is positive and increasing on $[0,1]$, the
smaller of the two roots is also $\lim_{t \to \infty} G_t(0)$. Therefore
by Theorem \ref{thm.forced}, the probability that a gate is forced in
$\alpha\log n$ steps is asymptotic to $\min(1,a/c)$.

In summary, for almost all Boolean networks,
almost all gates are $\alpha\log n$-weak if and only if $b + 2c
\le 1$, and almost all gates are forced in $\alpha\log n$ steps if and only if
$a/c \ge 1$. Since $a + b + c = 1$, $b + 2c \le 1$ is equivalent to $c \le a$.
Therefore both types of ordered behavior hold if and only if $a \ge c$.\footnote{
Articles \cite{l.crit} and \cite{l.thresh} contain proofs
that $a \ge c$ implies these kinds of ordered behavior;
it was conjectured in \cite{l.thresh} that
they fail when $a < c$.}

Kauffman performed extensive simulations on two classes of random networks
constructed from 2-argument Boolean functions. In the first class, all 16 of these
functions were equally likely to be assigned to a gate. In the second, no
constant functions were used, and the remaining 14 functions were equally likely.
In the first case, $a = 1/8$, $b = 3/4$, and $c = 1/8$, giving $\lambda = 1$ and
$g = 1$ as the only solution to Equation (\ref{eq.roots}). Therefore in this
case, almost all gates are weak and stable in $\alpha\log n$ steps.
But in the second case, $a = 0$, $b = 6/7$, and $c = 1/7$, giving $\lambda
= 8/7$ and $g = 0$ as the smaller root of (\ref{eq.roots}). Thus in this case,
a nontrivial fraction of the gates are $\alpha\log n$-strong and not forced in $\alpha\log n$
steps.
\section{Conclusions and Open Problems}
Our analysis for the case $a \ge c$ supports the experimental results for
networks of 2-input gates when all 16 2-argument functions are equally likely.
In fact, it gives stronger results than the conclusions of the experiments in
three senses. Kauffman's notion of weakness requires only that the network
should eventually return to the same limit cycle after a perturbation, but we
have shown that with high probability, within $\alpha\log n$ steps, the network
will return to exactly the same state it would be in without the perturbation.
Also, as mentioned earlier, forcing is a stronger condition than stability.
Lastly, the experiments indicated that almost all gates were weak and stabilized
for almost all inputs, while we have shown that almost all gates are weak and
forced for \emph{all} inputs.

On the other hand, there is a qualitative difference in the behavior of
random Boolean networks when $a < c$, and networks constructed from only the
14 nonconstant 2-argument functions belong to this category. However, this does
not necessarily contradict Kauffman's claim that these networks also display
ordered behavior since he stated only that, when perturbed they eventually
return to the same limit cycle, and eventually almost all gates stabilize.
It is possible that the effects of a perturbation vanish after $\alpha\log n$
steps, and most gates stabilize after $\alpha\log n$ steps.
Thus one open problem is to determine the long-term behavior of nets
where $a < c$ (or more generally, when $\lambda > 1$ or $g < 1$), to see if
the analysis agrees with the simulations.

We have not addressed the third of Kauffman's notions of order---the size of the
limit cycle, which Kauffman claims is of the order $\sqrt{n}$ for 2-input
networks. It has been shown that when $a > c$, 
not only is the average size of the limit cycle $O(\sqrt{n})$, it is bounded
by a constant with probability asymptotic to 1 \cite{l.crit}. However, when
$a = c$, the average size of the state cycle is superpolynomial in $n$
\cite{l.thresh}. To our knowledge, this is the only analytic result that
directly contradicts any of Kauffman's claims. The size of the limit cycle is
not known when $a < c$. We conjecture that it is superpolynomial in this case
also. More generally, it would be interesting to know if the size of the limit
cycle is determined by the $\lambda$ or $g$ parameters.

We have shown that one condition, $a \ge c$, imples both a large number of
weak gates and a large number of forced gates in networks of 2-input gates.
In the general case, two different conditions were used to characterize these
forms of order: $\lambda \le 1$ for weak gates, and $g = 1$ for forced gates.
Is there a single algebraic condition that characterizes both kinds of
order?

Other questions pertain to the effect of increasing the indegree of gates.
If we consider networks where each gate has $K$ inputs (using the uniform
distribution), then as mentioned in the Introduction, the simulations
indicate that when $K=2$, ordered behavior is very likely, but when
$K>2$, the networks tend to be disordered. We have described the results for
$K=2$ above. A similar analysis for $K>2$ remains to be done.
Using a different model of random Boolean network, B. Derrida and Y. Pomeau
\cite{dp} have provided evidence supporting the simulations. In their
version, at each step, each gate is randomly re-assigned its Boolean function and
its inputs. They referred to their model as the ``annealed" version and
Kauffman's as the ``quenched" version. They showed that, given any two
arbitrary initial states, as the two systems evolved over time, their Hamming
distance (the number of gates on which they differ) is approximated by
$c_Kn$ for some constant $c_K$ that depends on $K$. When $K=2$, $c_K = 0$,
but when $K>2$, $c_K>0$. Of course, when $K=2$, the quenched model behaves in
this way because almost all of the gates are forced. But it is not known whether
it holds for quenched models when $K > 2$, and the relationship between the
annealed and quenched models is not well understood.

Lastly, there is a network model that has some of the properties of both the
annealed and quenched models. Here, the gates and their connections are
fixed as in the quenched model, but at each step, a random collection of gates 
updates their states. In other words, the gates operate
asynchronously. As with the annealed model, an asynchronous network need not
enter a limit cycle, but the other notions of order are still meaningful,
and perhaps they can be studied productively.

\end{document}